# Orbital-selective charge transfer drives two-step negative thermal expansion structural transitions in PbTa$_2$Se$_4$

Peng Li[1,3,4*,†], Xiaohui Yang[2,5*], Wenhua Song[4], Zhefeng Lou[6], Tongrui Li[7], Zhengtai Liu[8], Zhu'an Xu[5,9], Zhuoyu Chen[1,4], Xiao Lin[6,†], and Yang Liu[3†]

[1]*Quantum Science Center of Guangdong–Hong Kong–Macao Greater Bay Area, Shenzhen 518045, China*

[2]*Department of Physics, China Jiliang University, Hangzhou 310018, China*

[3]*Center for Correlated Matter and School of Physics, Zhejiang University, Hangzhou 310058, China*

[4]*Department of Physics, Southern University of Science and Technology, Shenzhen 518055, China*

[5]*Zhejiang Province Key Laboratory of Quantum Technology and Device, Zhejiang University, Hangzhou 310027, China*

[6]*Key Laboratory for Quantum Materials of Zhejiang Province, Department of Physics, School of Science and Research Center for Industries of the Future, Westlake University, Hangzhou 310030, China*

[7]*Hefei National Laboratory, University of Science and Technology of China, Hefei 230088, China*

[8]*Shanghai Synchrotron Radiation Facility, Shanghai Advanced Research Institute, Chinese Academy of Sciences, Shanghai 201210, China*

[9]*School of Physics, Zhejiang University, Hangzhou 310058, China*

[*] These authors contributed equally.
[†]Emails: linxiao@westlake.edu.cn; yangliuphys@zju.edu.cn; lipeng@quantumsc.cn

**The negative thermal expansion (NTE) effect has been found generally combined with structural phase transitions. However, the charge and orbital freedoms of the NTE has not been well studied. This study employs angle-resolved photoemission spectroscopy and first-principles calculations to elucidate the charge and orbital kinetics of the anomalous two-step negative thermal expansion structural phase transitions in PbTa$_2$Se$_4$. As the temperature decreases, each transition undergoes a similar block-layer sliding, although the charge transfer behaviors differ significantly. During the first transition, charge is mainly transferred from the Pb 6$p_z$ orbital to an M-shaped band below the Fermi level, barely altering the Fermi surface. In contrast, the second transition involves modifications to both the Fermi surface and charge-transfer orbitals, with charge selectively transferred from Pb 6$p_x$/$p_y$ orbitals to Ta 5$d_{z^2}$ orbitals and a decrease of the Fermi pockets formed by Pb 6$p_x$/$p_y$ orbitals. Furthermore, a small pressure can easily tune the base structure phase among the three phases and the corresponding superconductivity. Therefore, our findings reveal that the orbital-selective charge transfer drives the unusual structure transition in PbTa$_2$Se$_4$, offering new insights into the NTE mechanisms and providing a unique window to study the pressure-tuned superconductivity in this metal-intercalated transition chalcogenides.**

**Introduction**

Intermetallic charge transfer frequently plays a pivotal role in triggering anomalous phenomena in transition metal chalcogenides (TMCs) [1-10], such as interface high-temperature superconductivity [2,11] and the negative thermal expansion (NTE) effect [5], due to simultaneous changes in valence states and significant variations in crystal structure. Stacking modulation and metal atom intercalation can effectively adjust interlayer charge distribution [3,12-15], thereby tuning interactions between various orders, such as those between charge density wave and superconductivity [16-22], and facilitating the emergence of novel phases, including topological superconductivity [18,23-27] and sliding ferroelectricity [28-31].

Angle-resolved photoemission spectroscopy (ARPES) is a powerful technique for probing electronic structures, revealing the valence band dispersions near the Fermi level as well as the core level characteristics well below it. The size of the measured Fermi pocket is directly connected with the material's carrier density. Specifically, the asymmetric line shape of core levels originates from interactions between suddenly created photoholes and conduction electrons, with the degree of asymmetry reflecting carrier density [32,33]. Furthermore, chemical shifts in core levels can indicate valence state changes of the corresponding element [34-36]. By analyzing both valence bands and core level variations, deeper insight into the charge transfer dynamics within the material can be obtained.

Recently, $PbTa_2Se_4$, with an onset superconducting transition temperature ($T_C$) of 1.8 K, has been found to exhibit anomalous two-step NTE phase transitions, caused by the relative sliding of its building block layers (Fig. 1a) [37]. Calculations also predict the existence of topological nodal lines in this material [37]. Raman spectroscopy experiments suggest that Ta plays a key role in these phase transitions [38]. Additionally, the phase transitions in $PbTa_2Se_4$ are also believed to be easily tunable by pressure, potentially modulating superconductivity, especially in comparison to $PbTaSe_2$ [23]. However, direct measurements of the temperature-dependent evolution of the electronic structure are still lacking, and the electronic origin of this unique two-step structure transition remains to be determined.

In this paper, we investigate the evolution of the electronic structure across the two-phase transitions using high-resolution ARPES. Combined with first-principles DFT calculations, we identify distinct driving forces behind the two-step phase transitions, revealing that a staged, orbital-selective charge transfer serves as the underlying mechanism. We further discuss the possible mechanism of the pressure-tuned superconductivity in this material through our transport measurements.

**Results and Discussion**

$PbTa_2Se_4$ crystallizes into a sandwiched structure where Pb layers are intercalated into a 3R-$TaSe_2$ framework, as illustrated in Fig. 1a. In this arrangement, each unit block comprises two $TaSe_2$ slabs separated by a single Pb layer [37]. Upon cooling from room temperature, the crystal structure of $PbTa_2Se_4$ undergoes two-step transitions, evolving from phase I to phase III. In each step, a relative sliding of the building block occurs, shifting the Pb–Se coordination from $PbSe_4$ tetrahedrons to $PbSe_2$ dumbbells [37]. Corresponding resistivity measurements (Fig. 1b) exhibit distinct thermal hysteresis in both transitions, confirming their first-order nature. During warming, the two transition temperatures are identified as $T_1 \approx 245$ K and $T_2 \approx 208$ K. The large energy-scale angle-integrated energy distribution curve (EDC) in Fig. 1c reveals sharp core levels of the corresponding elements, demonstrating the high crystallinity of the measured sample. The overlapping black curve represents the Pb core level of the mixed termination, where both Pb-termination and $TaSe_2$-termination contribute to the intensity and spectral line shape. The single-peak red curve corresponds to the $TaSe_2$-termination observed in the subsequent data analysis [24].

Figure 1d schematically depicts the bulk and surface Brillouin Zones (BZ).

To elucidate the mechanisms underlying the two distinct transitions, a deeper investigation of the electronic structures is required. Figure 2 displays the valence band electronic structures of the three PbTa$_2$Se$_4$ phases measured at 20 K, 225 K and 270 K. In the Fermi surface (FS) map of phase III, at least five pockets (labeled as $\alpha$, $\beta$, $\gamma$, $\delta$ and $\varepsilon$) are identified (Fig. 2a). The temperature-dependent evolution of band dispersions along the high symmetry cut $\overline{\Gamma K}$ (marked by the red dashed line in Fig. 2a) is illustrated in Figs. 2d-2f. The $\delta$ band is consisted of two nearly degenerate electron bands. The most prominent distinction between phases II and I lies in the emergence of an M-shaped band (the black curves in Figs. 2d and 2e) below the Fermi level at the $\overline{\Gamma}$ point. In phase I, the M-shaped band bottom seems touch the Fermi level and forms a weak and small electron pocket. This M-shaped band shifts to higher binding energy in phase III with much higher spectral intensity compared to phase II. The FS maps and majority of band dispersions in phases II and I exhibit strong similarities, while they differ markedly from those of phase III. The $\delta$ band abruptly shifts to lower binding energy and the $\gamma$ band is well formed in phase III. Figure 4b summarizes the band structures of all three phases, with extended temperature evolution data provided in Fig. S1 and Fig. S2.

Additionally, $k_z$-$k_y$ maps (Figs. 2g-i) along the red dashed line in Fig. 2a at the Fermi energy (E$_F$) exhibit straight line-like features along the $k_z$ direction across all three phases, indicating quasi-two-dimensional electronic structures near the Fermi level. From our photoenergy-dependent measurements shown in Figs. S3-S5, where there is no obvious intensity change of the M-shaped band along the $k_z$ direction, we could conclude that the significant intensity increase of the M-shaped band from phase II to phase III is not $k_z$-related intensity variation but originates from intrinsic electronic properties.

Figure 3a illustrates the orbital occupancy of Pb-6$p$, Ta-5$d$ and Se-4$p$ in phase III derived from DFT calculations, offering valuable insight for interpreting temperature-dependent ARPES data and resolving orbital-specific changes. The calculations reveal that the $\delta$ electron band at the K point is dominated by in-plane Pb 6$p_x$/$p_y$ orbitals, the $\alpha$ band (at least two bands in calculations which cannot be resolved by ARPES) predominantly originates formed the Ta $5d_{z^2}$ orbital. The M-shaped band emerges from hybridization involving out-of-plane Pb 6$p_z$, Ta $5d_{z^2}$ and Se 4$p_z$ orbitals. Notably, most Se-derived orbitals reside predominantly below the Fermi level, while the Fermi surface is primarily governed by contributions from Pb and Ta orbitals. The $\beta$ band observed in ARPES spectra corresponds to a surface state, which is characteristic of the TaSe$_2$-terminated surface (Figs. 3b-d) but absent in the Pb-terminated configuration (Figs. 3e-g). In calculations, the M-shaped band bottom also shifts to higher binding energy from phase I to phase III, giving qualitative guidance for experimental results.

The geometric parameters of the Fermi surface pockets directly correlate with the carrier density. By measuring the Fermi momentum ($k_F$) of each pocket at different temperatures, we can track the temperature evolution of the material's carrier density. Figure 4a shows the extracted momentum distribution curves (MDCs) at the Fermi level along the cuts shown in Figs. 2d-f, with constant intensity shifts and dashed lines highlighting peak positions. The valence band structures of the three phases, as summarized experimentally in Fig. 4b, provide an intuitive illustration of the band changes. Systematic analysis in Fig. 4c reveals that the carrier density remains unchanged between phase I and phase II, while a sudden reduction in carrier density occurs from phase II to phase III, particularly for the $\alpha$ (Ta $5d_{z^2}$) and $\delta$ (Pb 6$p_z$) bands.

The core level shift of the element reflects the charge transfer [34,36], while the asymmetry of the line shape resulted from the interaction of the suddenly created potential of the photohole with the conduction electrons can indirectly indicate the carrier density. The asymmetric core line of a metal can be fitted using the following Doniach-Sunjic equation [33,39-43]:

$$I(\epsilon) = \frac{\Gamma(1-\Lambda)}{(\epsilon^2+\tau^2)^{(1-\Lambda)/2}} \cos\left[\frac{1}{2}\pi\Lambda + \theta(\epsilon)\right] \quad (1)$$

where $\theta(\epsilon) = (1-\Lambda)tan^{-1}(\epsilon/\tau)$, $\Gamma$ denotes the Gamma function, and $\tau$ represents the lifetime of the core-hole state. The asymmetry coefficient $\Lambda$ gives a skewed line shape tailing towards higher binding energy. A fit of Eq. (1) therefore provides a rough estimate of $\Lambda$. Additionally, the energy position can be also extracted from this analysis. Another possible explanation for the asymmetric line shape is the presence of satellite peaks, potentially arising from core levels of a mixed-terminated surface. However, we rule out this possibility based on Fig. 1c and Fig. S6, which presents the line shape of a mixed-terminated surface from a different measured sample. In that case, the Pb 5d core levels associated with different termination types are clearly resolved as separate, distinct peaks. Thus, we confidently conclude that the asymmetry in the core-level line shape well-described by the Doniach-Sunjic equation is not from satellite peaks.

Figure 4d displays the Pb $5d_{5/2}$ and Pb $5d_{3/2}$ core levels at various temperatures from 275 K to 70 K, with constant intensity shifts. The complete overlap between raw data and red-dashed fitting curve confirms accurate spectral deconvolution (details in Fig. S7). Figures 4e and 4f plot the energy positions and asymmetry coefficient ($\Lambda$) of the Pb $5d_{3/2}$ core level extracted from the fittings. During the first transition (phase I to II) at $T_1$, the energy position shifts towards higher binding energies, suggesting electron transfer from Pb to TaSe$_2$ layers, concomitant with Pb valence enhancement. Crucially, this charge redistribution occurs without $\Lambda$ variation (remaining constant), demonstrating preserved carrier density through phases II, which is consistent with the constant Fermi momentum ($k_F$) results as shown in Figs. 4b and 4c. The small electron band at the $\bar{\Gamma}$ point in phase I contributes little to carrier density. In stark contrast, the second transition (phase II to III) at $T_2$ manifests concurrent discontinuities: the binding energy shifts with different velocity and $\Lambda$ undergoes a step-like increase at the transition temperature. Notably, this step-like shift in $\Lambda$ aligns well with the results in Fig. 4c, where the Fermi momentum of the $\delta$ and $\alpha$ bands suddenly decrease, further supporting the sudden change in carrier density. Although the well-resolved $\gamma$ band emerges in phase III, it would persist in phases II and I, consistent with calculations, but with rather weak intensity. This dichotomy explicitly resolves the electronic origins of both transitions: the first driven by more localized charge transfer and the second governed by collective carrier density reorganization.

Having identified the charge transfer and carrier density variations during both phase transitions, it becomes imperative to jointly consider these factors for elucidating the underlying mechanisms. In the $T_1$ transition, the loss of electrons from Pb does not contribute to the Fermi surface but rather form the M-shaped band (bonding band) below the Fermi level. This occurs due to enhanced hybridization among Pb $6p_z$, Ta $5d_{z2}$ and Se $5p_z$ orbitals in PbSe$_2$ dumbbells [37]. Meanwhile, the electron bands at the $\bar{K}$ point formed by Pb $6p_x/p_y$ orbitals remain unchanged in both phases I and II, indicating that the charge transfer at $T_1$ localized at Pb $6p_z$ orbital. In Fig. 4b, the continuous electron transfer from Pb $6p_z$ orbital into the TaSe$_2$ layer persist throughout phase II until the $T_2$ transition abruptly modifies the slope of the energy position curve. As for the $T_2$ transition, while the intensity of the M-shaped band progressively enhances with cooling (details in Fig. S8), more pronounced modifications emerge in Fermi-level-crossing bands. Particularly, the $\alpha$ and $\delta$ bands shrink in phase III. According to the orbital characters illustrated in Fig. 3a, these changes can be attributed to charge transfer from Pb $6p_x/p_y$ orbitals to Ta $5d_{z2}$, $5d_{xy}$ and $5d_{x2-y2}$ orbitals. It is also worth noting that the $2k_F$ of the $\beta$ band remains constant in all phases likely due to its surface state origin. Therefore, the charge transfer does not merely result in a simple rigid shift of the Fermi level but instead represents a more complex process. Recently, the nonrigid shift of band dispersions has also been found in Cu-intercalated 2H-TaSe$_2$ [44].

Our previous results show that PbTa$_2$Se$_4$ exhibits an onset $T_C$ of 1.8 K, significantly higher than pure 2H-TaSe$_2$ ($T_C$ ~ 0.14 K) [45], yet approximately half the $T_C$ of PbTaSe$_2$ ($T_C$ ~ 3.8 K) [18,24-26,46,47] and other optimized metal-intercalated TaSe$_2$ compounds [17,20,48], such as Pd$_{0.08}$TaSe$_2$ ($T_C$ ~ 2.8 K) [20]. Notably, PbTaSe$_2$ also undergoes a similar NTE transition at 405 K, where a weak pressure of 0.25 GPa drives the high-temperature phase into the ground state with tuned superconductivity and structure phase. Previous theoretical studies on PbTaSe$_2$ have indicated that the in-plane orbitals (Pb $6p_x/p_y$, Ta $5d_{x2-y2}/d_{xy}$) near the $\overline{K}$ point exhibit the strongest electron-phonon coupling and host the largest superconducting gap [47]. Therefore, the superconductivity and structure phase of PbTa$_2$Se$_4$ could be readily tuned by small pressure variations, and our ARPES studies provide unique insights into the electronic origin of the potential pressure-tuned superconductivity. Based on our ARPES studies, phase I and phase II present similar Fermi surface characteristics, suggesting comparable superconducting $T_C$ values with ground structure of these two phases. In contrast, phase III may exhibit a distinct $T_C$ due to modifications in the in-plane orbitals (Pb $6p_x/p_y$, Ta $5d_{x2-y2}/d_{xy}$) near the $\overline{K}$ point. Fig. 5 presents the pressure-tuned superconductivity of PbTa$_2$Se$_4$, in which a small pressure of 0.2 GPa can absolutely suppress the two-step structure transition and enhance the onset $T_C$ to be 2.8 K at 0.3 GPa. In Fig. 5c, the superconducting transition temperatures with ground structure of phases II or I are significantly larger than those with ground structure of phase III. The enhancement of superconductivity in structures of phases II and I strongly suggest the key role of the in-plane orbitals (Pb $6p_x/p_y$, Ta $5d_{x2-y2}/d_{xy}$) in superconducting pairing in PbTa$_2$Se$_4$.

In summary, we performed a detailed ARPES investigation into the orbital characteristics during the two-step phase transitions in PbTa$_2$Se$_4$, identifying distinct orbital-selective charge transfer mechanisms through analysis of Fermi surface evolution and Pb core-level asymmetry with temperature. The first transition (phase I to phase II) involves selective charge transfer from the Pb $6p_z$ orbital to TaSe$_2$ layer, with the M-shaped band forming below the Fermi level due to enhanced hybridization between Pb $6p_z$, Ta $5d_{z2}$ and Se $4p_z$ orbitals. In contrast, the second transition (phase II to phase III) features charge transfer from Pb $6p_x/p_y$ orbitals to conductive Ta $5d_{z2}$ and Ta $5d_{x2-y2}/5d_{xy}$ orbitals, inducing significant Fermi surface modifications and carrier density changes. A small pressure can readily tune the superconductivity, with the enhanced superconductivity linked to the ground-state structure of phase II or I under pressure. Our findings not only highlight the crucial role of orbital-selective charge transfer in driving the two-step NTE phase transitions in PbTa$_2$Se$_4$ but also provide valuable insights for understanding the potential pressure-tuned superconductivity mechanism in this compound and other metal-intercalated TMCs.

**Methods**

**ARPES experiments**

High quality PbTa$_2$Se$_4$ single crystals were synthesized by the chemical vapor transport (CVT) method described in Ref. [37]. Synchrotron-ARPES measurements were carried out using synchrotron light sources at BL 03U of Shanghai Synchrotron Radiation Facility (SSRF) in China.

The overall energy resolution was set to be better than 10 meV at 110 eV photon energy and the angular resolution is ~ 0.2 degree for the measurements. The crystals were cleaved *in-situ* and measured with a base pressure better than $6 \times 10^{-11}$ Torr. All the data presented in this paper were taken within a few hours after cleavage ensuring the results were not affected by aging effect.

**Band calculations**
Electronic structure calculations were performed using density functional theory (DFT) with a plane wave basis projected augmented wave, as implemented in the Vienna ab-initio simulation package (VASP) [49]. The Perdew-Burke-Ernzerhof (PBE) approximation was used as the exchange-correlation potential [50]. Wannier functions were constructed by projecting Bloch states onto Pb-6*p*, Ta-5*d*, and Se-4*p* orbitals through WANNIER90 [51,52]. An energy cutoff of 300 eV and $8 \times 8 \times 8$ Γ-centered k-mesh were employed in the calculation. Fermi levels were shifted up by 70 meV to match the ARPES spectra.

**Hydrostatic pressure measurements**
Samples were loaded into a piston-type pressure cell and the actual pressure was determined by measuring the superconducting transition temperatures of Pb. Daphne 7373 oil was applied as the pressure transmission media. For the data under different pressures, the same contacts were used throughout the measurements such that the geometric errors in the contact size were identical for different runs.

**Data Availability**
All data needed to evaluate the conclusion in the paper are present in the paper and/or the Supplementary information.

**Acknowledgements**
We would like to thank Dr. Zhicheng Jiang for help in the synchrotron ARPES measurements. This work is supported by the Guangdong Provincial Quantum Science Strategic Initiative (Grant No. GDZX2401004 (P. L.)), National Natural Science Foundation of China (Grants No. 12304168 (X. H. Y.)), Zhejiang Provincial Natural Science Foundation of China (Grant No. LQ23A040009 (X. H. Y.)), Zhejiang Provincial Natural Science Foundation of China for Distinguished Yong Scholars (Grant No. LR23A04001 (X. L.)) and the National Key R&D Program of China (Grant No. 2022YFA1402200 (Y. L)).

**Author contributions**
P. L. and X. H. Y. contribute equally to this work. P. L. and X. H. Y. conceived the experiments. P. L. and X. H. Y. carried out ARPES measurements with assistance of Z. T. L., T. R. L. and Z. F. L. performed the DFT calculations. P. L. and W. H. S. analyze the ARPES data, X. H. Y and X. L. synthesized and characterized bulk single crystals. X. L. conducted the pressure-tuned transport measurements. X. H. Y., Y. L. and P. L. wrote the manuscript. All authors contributed to the scientific planning and discussions.

**Competing Interests**
The authors declare no competing interest.

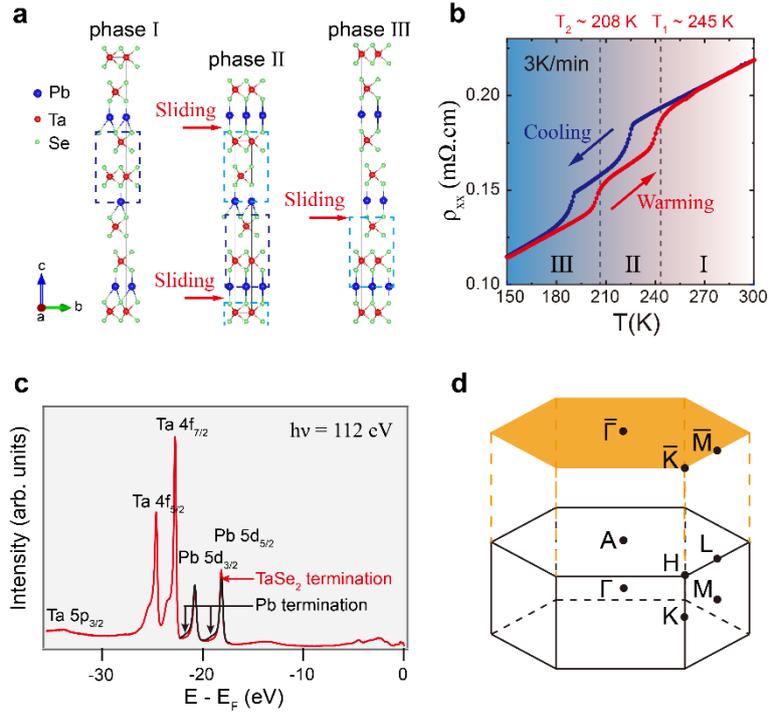

**Fig. 1 Basic information of PbTa$_2$Se$_4$. a** The crystal structures of three phases: phase I (high temperature), phase II (intermedium temperature) and phase III (low temperature). The deep blue dashed indicates the basic block layer, being consisted of one 2H-TaSe$_2$ layer and one Pb atom layer. From phase I to phase III, the crystal structures can be viewed as the results of two-step sliding of the block layer in one unite cell. **b** The resistivity measurements of the measured sample, the huge thermal hypothesis during the cooling and warming processes indicates the first order nature of the two transitions. The two transition temperatures during the warming process are identified as $T_1 \sim$ 245 K (from phase I to phase II) and $T_2 \sim$ 208 K (from phase II to phase III). **c** The angle-integrated energy distribution curve using photon energy of 112 eV. **d** The schematic of the bulk and surface Brillouin Zones.

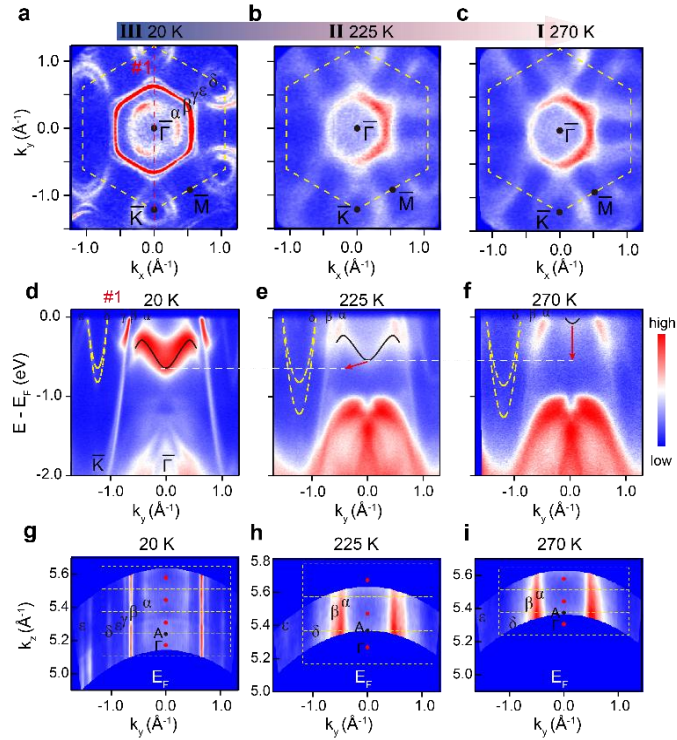

**Fig. 2 Electronic structures at three phases. a-c** The Fermi surface maps measured at 20 K, 225 K and 270 K, respectively, using photon energy of 110 eV. The yellow dashed line indicates the edge of the surface BZ and the red dashed line indicates the high symmetry line along $\overline{\Gamma K}$. The pockets in the Fermi surface map of phase III are labeled as as $\alpha$, $\beta$, $\gamma$, $\delta$ and $\varepsilon$, respectively. **d-f** The corresponding ARPES spectra of the three phases along the red dashed line in **a**. The yellow dashed lines indicate the electron bands at the $\overline{K}$ point. **g-i** The $k_z$-$k_y$ maps at the Fermi level by photoenergy-dependent measurements of the three phases along the red dashed line in **a**. The yellow dashed rectangular is the bulk BZs along the $k_z$ direction. The color bar shows the ARPES spectra intensity.

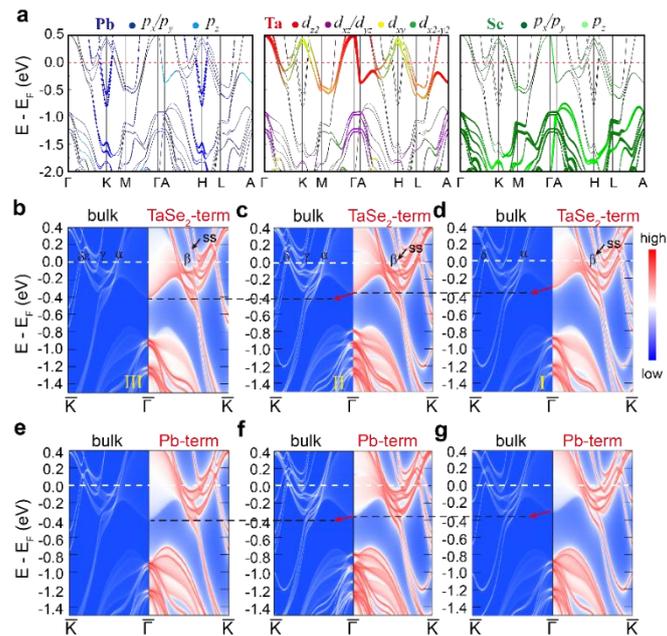

**Fig. 3 Orbital characteristics and band evolution across three phases, with calculated band structures for different surface terminations. a** Orbital occupations of Pb 6$p$ ($p_x$, $p_y$), Ta 5$d$ ($d_{z^2}$, $d_{xz}/d_{yz}$, $d_{xy}$, $d_{x^2-y^2}$) and Se 4$p$ ($p_x/p_y$, $p_z$) in phase III. **b-d** Comparison of Se-terminated surface band structures along $\overline{\Gamma K}$ with calculated bulk bands, identifying the $\beta$ band as a surface state. **e-f** Pb-terminated surface band structures along $\overline{\Gamma K}$ compared to bulk bands, showing no correspondence of the observed $\beta$ band. The color bar shows the band intensity.

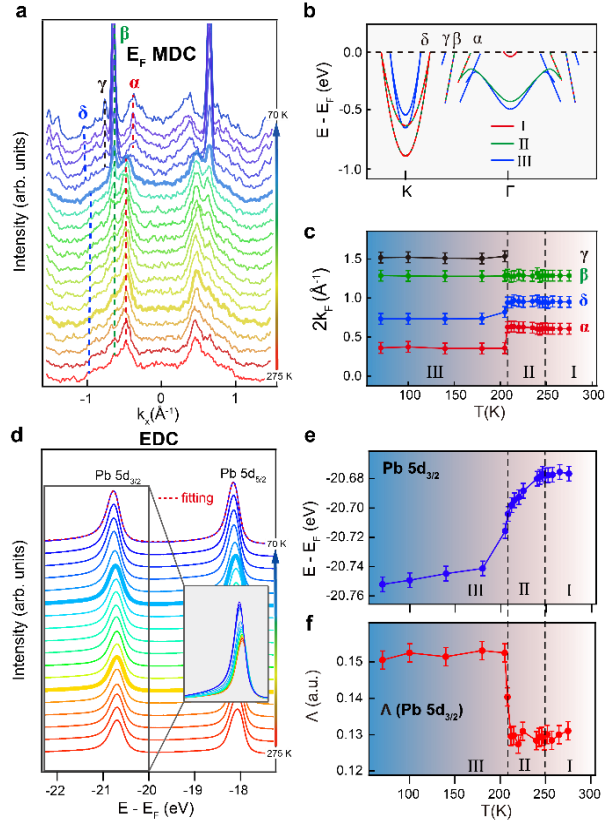

**Fig. 4 Orbital-dependent charge transfer and carrier density changes with temperature. a** Temperature evolution of the momentum distribution curves at Fermi level along $\overline{\Gamma K}$. The red, green, black and blue dashed lines are guides for eyes for the peak positions of the $\alpha$, $\beta$, $\gamma$ and $\delta$ bands, respectively. **b** The schematic of the bands in three phases along $\overline{\Gamma K}$ extracted from experiments. **c** Twice of the Fermi vector ($2k_F$) of the $\alpha$, $\beta$, $\gamma$ and $\delta$ bands at different temperatures. **d**. A series of energy distribution curves from 275 K to 70 K with constant intensity shifts. The red dashed line is the fitting curve. The inset shows the Pb 5$d_{3/2}$ peaks without intensity shifts, well showing the asymmetry line shape and temperature evolutions. **e** The peak position of Pb 5$d_{3/2}$ with the function of temperature. **f** The corresponding asymmetry coefficient Λ with the function of temperature. The black dashed lines indicate the position of the two transition temperatures.

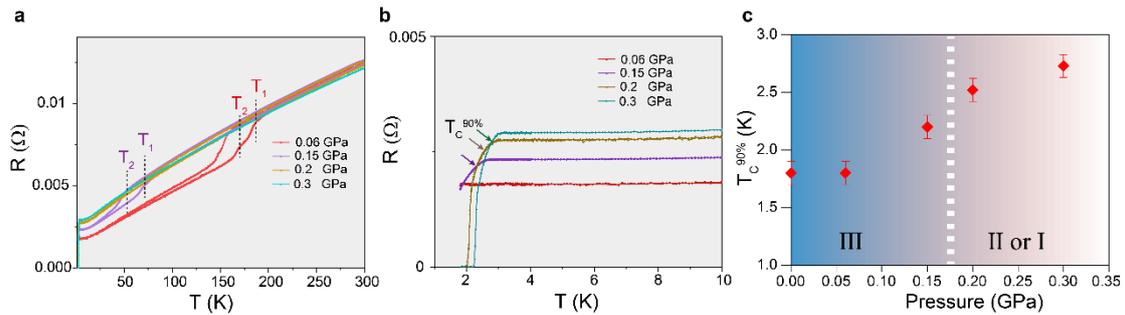

**Fig. 5 Pressure-tuned structure phase transition and superconductivity of PbTa$_2$Se$_4$. a** Temperature dependent resistivity curves of PbTa$_2$Se$_4$ under small pressures (0.06, 0.15, 0.2, and 0.3 GPa). **b** Zoom-in view of the resistivity curves near superconducting transition temperatures. **c** Summary of the superconducting onset temperatures under pressure for the corresponding structure phases. The onset temperature is determined at the 90% of the normal state values. The value of ambient pressure is extracted from ref. [37].